\documentclass[12 pt,twoside,aps,prd,amsmath,amssymb,
tightenlines,showpacs,showkeys,eqsecnum]{revtex4}
\usepackage{mathptmx}
\usepackage{bm}
\usepackage{graphicx}
\usepackage{hyperref}
\renewcommand{\cal}{\mathcal}

\newcommand {\pr}{\partial}

\newcommand {\cL}{\cal L}
\newcommand {\cD}{\cal D}

\def \myfigures #1#2#3#4#5#6#7#8
{\begin{figure}[ht]
    \begin{center}
        \includegraphics[width=#2 \textwidth]{#1.eps}
        \hfill
        \includegraphics[width=#6 \textwidth]{#5.eps}
        \parbox[t]{#4\textwidth}{\caption {#3}\label{#1}}
        \hfill
        \parbox[t]{#8\textwidth}{\caption {#7}\label{#5}}
    \end{center}
\end{figure} }

\def\myfigure #1#2#3#4
{\begin{figure}[ht]\begin{center}
\includegraphics[width=#2 \textwidth]{#1.eps}
\parbox[t]{#4\textwidth}{\caption{#3}\label{#1}}
\end{center}\end{figure}}

\date{\today}
\begin{document}
\title{Electromagnetic field with induced massive term: Case with spinor field}
\author{Yu.P. Rybakov, G.N. Shikin and Yu.A. Popov}
\affiliation{Department of Theoretical Physics\\
Peoples' Friendship University of Russia\\
117198 Moscow, Russia} \email{soliton4@mail.ru}

\author{Bijan Saha}
\affiliation{Laboratory of Information Technologies\\
Joint Institute for Nuclear Research, Dubna\\
141980 Dubna, Moscow region, Russia} \email{bijan@jinr.ru}
\homepage{http://bijansaha.narod.ru}

\begin{abstract}
We consider an interacting system of spinor and electromagnetic
field, explicitly depending on the electromagnetic potentials, i.e.,
interaction with broken gauge invariance. The Lagrangian for
interaction is chosen in such a way that the electromagnetic field
equation acquires an additional term, which in some cases is
proportional to the vector potential of the electromagnetic field.
This equation can be interpreted as the equation of motion of photon
with induced non-trivial rest-mass. This system of interacting
spinor and scalar fields is considered within the scope of Bianchi
type-I (BI) cosmological model. It is shown that, as a result of
interaction the electromagnetic field vanishes at $t \to \infty$ and
the isotropization process of the expansion takes place.
\end{abstract}

\keywords{electromagnetic field, spinor field, Bianchi type I (BI)
model, photon mass}

\pacs{03.65.Pm and 04.20.Ha}

\maketitle

\bigskip


\section{Introduction}

Though the triumphs of Maxwellian electromagnetism and quantum
electrodynamics set the rest mass of photon to be trivial, the
hypothesis of possible nontrivial photon mass has long been
discussed in the literature \cite{gold,froome,taylor,broglie}. The
modern experimental data do not contradict this hypothesis
\cite{rosen,williams,crandal,chernikov,schaefer,fishbach,davis,lakes,luo}.
So it is interesting to consider some additional arguments for or
against this hypothesis. As one of such arguments can serve
experimental data of modern observational cosmology, which witnesses
the isotropy of the Universe. It is interesting to combine this fact
with the description of matter by means of system of interacting
fields including the electromagnetic one. In a recent paper
\cite{shikin1} we considered one of the simplest systems comprising
with mass-less scalar and electromagnetic  fields and study the
influence of such interaction on the expansion of the Universe in
the asymptotic region. In that paper it was shown that if one
consider only electromagnetic field, the two of the three spatial
components vector potential are either constant or zero and the
space-time in this case does not allow isotropization.

In the recent years system with nonlinear spinor field was
extensively studied in different cosmological models and it was
shown that the nonlinear spinor field plays very important role in
(i) isotropization of initially anisotropic space-time, (ii)
formation of singularity free cosmological solutions, and (iii)
explaining late-time acceleration
\cite{shikin2,saha1,saha2,saha3,saha4}. In connection with this in
this paper we consider an system of spinor and electromagnetic
fields within the scope of a Bianchi type-I cosmological model and
examine the influence of such interaction on the expansion of the
Universe in the asymptotic region.

\section{Basic equations and their general solutions}

We choose the Lagrangian of the interaction electromagnetic and
spinor fields within the framework of a BI cosmological
gravitational field in the form
\begin{equation}
{\cL} = \frac{R}{2\kappa} - \frac{1}{4} F_{\mu\nu}F^{\mu\nu} +
\frac{i}{2} \biggl[\bar \psi \gamma^{\mu} \nabla_{\mu} \psi-
\nabla_{\mu} \bar \psi \gamma^{\mu} \psi \biggr] - m \bar \psi \psi
+ \frac{1}{2} K(I) {\cD}(S),\ \label{lag}
\end{equation}
with $I = A_\lambda A^\lambda$ and $S = \bar \psi \psi$. We consider
the BI metric in the form
\begin{equation}
ds^2 = e^{2\alpha} dt^2 - e^{2\beta_1} dx^2 - e^{2\beta_2} dy^2 -
e^{2\beta_3} dz^2. \label{BI}
\end{equation}
The metric functions $\alpha, \beta_1, \beta_2, \beta_3$ depend on
$t$ only and obey the coordinate condition
\begin{equation}
\alpha = \beta_1 + \beta_2 + \beta_3. \label{cc}
\end{equation}

Written in the form
\begin{equation}
R_{\mu}^{\nu} = -\kappa \Bigl(T_{\mu}^{\nu} - \frac{1}{2}
\delta_{\mu}^{\nu} T \Bigr), \label{ee}
\end{equation}
the Einstein equations corresponding to the metric \eqref{BI} in
account of \eqref{cc} read
\begin{subequations}
\label{BID}
\begin{eqnarray}
e^{-2\alpha} \Bigl(\ddot \alpha - \dot \alpha^2 + \dot \beta_1^2 +
\dot \beta_2^2 + + \dot \beta_3^2 \Bigr) &=&  - \kappa
\Bigl(T_{0}^{0} - \frac{1}{2} T \Bigr),\label{00}\\
e^{-2\alpha} \ddot \beta_1 &=&  - \kappa
\Bigl(T_{1}^{1} - \frac{1}{2} T \Bigr),\label{11}\\
e^{-2\alpha} \ddot \beta_2 &=&  - \kappa
\Bigl(T_{2}^{2} - \frac{1}{2} T \Bigr),\label{22}\\
e^{-2\alpha} \ddot \beta_3 &=&  - \kappa \Bigl(T_{3}^{3} -
\frac{1}{2} T \Bigr),\label{33}
\end{eqnarray}
\end{subequations}

where over dot means differentiation with respect to $t$ and
$T_{\nu}^{\mu}$ is the energy-momentum tensor of the material field.

Variation of \eqref{lag} with respect to electromagnetic field gives
\begin{equation}
\frac{1}{\sqrt{-g}} \frac{\pr}{\pr x^\nu} \Bigl(\sqrt{-g}
F^{\mu\nu}\Bigr) -  {\cD}(S) \frac{dK}{dI} A^{\mu} = 0. \label{emf}
\end{equation}
The spinor field equation corresponding to the metric \eqref{lag}
has the form
\begin{subequations}
\begin{eqnarray}
i\gamma^{\mu} \nabla_{\mu} \psi - m \psi + \frac{1}{2} K(I)
\frac{d{\cD}}{dS} \psi &=& 0, \label{psi}\\
i \nabla_{\mu} \bar \psi \gamma^{\mu} + m \bar \psi - \frac{1}{2}
K(I) \frac{d{\cD}}{dS} \bar \psi &=& 0. \label{bpsi}
\end{eqnarray}
\end{subequations}
The energy-momentum tensor of the interacting matters fields has the
form
\begin{eqnarray}
T_\mu^\nu &=& -  F_{\mu \eta} F^{\nu\eta} + \frac{i}{4} g^{\nu \rho}
\Bigl(\bar \psi \gamma_{\rho} \nabla_{\mu} \psi + \bar \psi
\gamma_{\mu} \nabla_{\rho} \psi- \nabla_{\mu} \bar \psi
\gamma_{\rho} \psi - \nabla_{\rho} \bar \psi \gamma_{\mu} \psi
\Bigr) \nonumber\\ &+&  {\cD}(S) \frac{dK}{dI} A_{\mu}A^{\nu} -
\delta_\mu^\nu {\cL}. \label{emt}
\end{eqnarray}
We consider the case when the electromagnetic and scalar fields are
the functions of $t$ only.  Taking this in mind we choose the vector
potential in the following way:
\begin{equation}
A_{\mu} = \bigl(0,\,A_1(t),\,A_2(t),\,A_3(t)\bigr). \label{vpot}
\end{equation}
In this case the electromagnetic field tensor $F^{\mu \nu}$ has only
three non-vanishing components, namely
\begin{equation}
F_{01} = \dot A_1, \quad F_{02} = \dot A_2, \quad F_{03} = \dot A_3.
\label{emtensor}
\end{equation}
On account of \eqref{vpot} and \eqref{emtensor} we now have
\begin{eqnarray}
I &=& A_\lambda A^\lambda = - A_1^2 e^{-2\beta_1} - A_2^2
e^{-2\beta_2} - A_3^2 e^{-2\beta_3}, \label{eminv}\\
F_{\mu\nu}F^{\mu\nu} &=& - 2 e^{-2\alpha} \bigl(\dot A_1^2
e^{-2\beta_1} + \dot A_2^2  e^{-2\beta_2} + \dot A_3^2e^{-2\beta_3}
\bigr). \label{fmn}
\end{eqnarray}
For electromagnetic field from \eqref{emf} we find
\begin{subequations}
\label{emf123}
\begin{eqnarray}
\frac{d}{dt}\Bigl(\dot A_1 e^{-2\beta_1}\Bigr)  + {\cD}(S)
\frac{dK}{dI} e^{2\alpha - 2\beta_1} A_1 &=& 0, \label{emf1}\\
\frac{d}{dt}\Bigl(\dot A_2 e^{-2\beta_2}\Bigr)  + {\cD}(S)
\frac{dK}{dI} e^{2\alpha - 2\beta_2} A_2 &=& 0, \label{emf2}\\
\frac{d}{dt}\Bigl(\dot A_3 e^{-2\beta_3}\Bigr) + {\cD}(S)
\frac{dK}{dI} e^{2\alpha - 2\beta_3} A_3 &=& 0. \label{emf3}
\end{eqnarray}
\end{subequations}

Let us now go back to the spinor field equation. The spinor field
equation for the metric \eqref{BI} takes the form
\begin{equation}
i e^{-\alpha} \bar \gamma_0 \Bigl(\partial_t + \frac{\dot \alpha}{2}
\Bigr) \psi - m \psi + \frac{1}{2} K(I) \frac{d {\cD}}{dS} \psi = 0,
\label{spinorBI}
\end{equation}
where $\bar \gamma_\mu$ is the Dirac matrices for flat space-time.
Equation \eqref{spinorBI} can be rewritten as
\begin{subequations}
\label{spinsys}
\begin{eqnarray}
\dot \psi_a + \frac{\dot \alpha}{2} \psi_a + i e^{\alpha} \Bigl(m -
\frac{1}{2} K(I) \frac{d{\cD}}{dS}\Bigr) \psi_a &=& 0, \quad a
=1,\,2, \\
\dot \psi_b + \frac{\dot \alpha}{2} \psi_b - i e^{\alpha} \Bigl(m -
\frac{1}{2} K(I) \frac{d{\cD}}{dS}\Bigr) \psi_a &=& 0, \quad b
=3,\,4.
\end{eqnarray}
\end{subequations}
Solving \eqref{spinsys} we find the following components of the
spinor fields
\begin{subequations}
\label{spinsyscomp}
\begin{eqnarray}
\psi_a &=& C_a \exp{\Bigl[-\frac{\alpha}{2} - i \int \Bigl(m -
\frac{1}{2} K(I) \frac{d{\cD}}{dS}\Bigr) e^{\alpha} dt\Bigr]},\quad
C_a = {\rm const.} \\
\psi_b &=& C_b \exp{\Bigl[-\frac{\alpha}{2} + i \int \Bigl(m -
\frac{1}{2} K(I) \frac{d{\cD}}{dS}\Bigr) e^{\alpha} dt\Bigr]},\quad
C_b = {\rm const.}
\end{eqnarray}
\end{subequations}
From \eqref{spinsyscomp} we also find
\begin{equation}
S = S_0 e^{- \alpha}. \label{salpha}
\end{equation}

Finally, let us solve the Einstein equations. In doing so, let us
first write the non trivial components of the energy momentum tensor
of material fields. In view of \eqref{spinsys} from \eqref{emt} we
find
\begin{subequations}
\label{emtcomp}
\begin{eqnarray}
T_0^0 &=& mS - \frac{1}{2} K(I) {\cD}(S) + \frac{1}{2}\Bigl(\dot
A_1^2 e^{-2\beta_1} + \dot A_2^2 e^{-2\beta_2} + \dot
A_3^2 e^{-2\beta_3}\Bigr) e^{-2\alpha}, \label{00emt}\\
T_1^1 &=&  \frac{1}{2}K(I)\Bigl(\frac{d{\cD}}{dS} S - {\cD}(S)\Bigr)
+\frac{1}{2} \Bigl(\dot A_1^2 e^{-2\beta_1} - \dot A_2^2
e^{-2\beta_2} - \dot A_3^2 e^{-2\beta_3}\Bigr)e^{-2\alpha} \nonumber
\\ &-& {\cD} \frac{dK}{dI}A_1^2 e^{-2\beta_1}, \label{11emt}\\
T_2^2 &=&  \frac{1}{2}K(I)\Bigl(\frac{d{\cD}}{dS} S - {\cD}(S)\Bigr)
+\frac{1}{2} \Bigl(\dot A_2^2 e^{-2\beta_2} - \dot A_3^2
e^{-2\beta_3} - \dot A_1^2 e^{-2\beta_1}\Bigr) e^{-2\alpha}
\nonumber \\ &-& {\cD} \frac{dK}{dI} A_2^2 e^{-2\beta_2}, \label{22emt}\\
T_3^3 &=& \frac{1}{2}K(I)\Bigl(\frac{d{\cD}}{dS} S - {\cD}(S)\Bigr)
+\frac{1}{2}\Bigl(\dot A_3^2 e^{-2\beta_3} - \dot A_1^2
e^{-2\beta_1} - \dot A_2^2 e^{-2\beta_2}\Bigr) e^{-2\alpha}
\nonumber \\ &-& {\cD}
\frac{dK}{dI} A_3^2 e^{-2\beta_3}, \label{33emt}\\
T_1^2 &=& \Bigl( \dot A_1 \dot A_2 e^{-2 \alpha}  - {\cD}
\frac{dK}{dI} A_1 A_2 \Bigr) e^{- 2\beta_1}. \label{12emt}\\
T_2^3 &=& \Bigl( \dot A_2 \dot A_3 e^{-2 \alpha}  - {\cD}
\frac{dK}{dI} A_2 A_3 \Bigr) e^{- 2\beta_2}. \label{23emt}\\
T_3^1 &=& \Bigl( \dot A_3 \dot A_1 e^{-2 \alpha}  - {\cD}
\frac{dK}{dI} A_3 A_1 \Bigr) e^{- 2\beta_3}. \label{31emt}
\end{eqnarray}
\end{subequations}
From \eqref{emtcomp} one also finds
\begin{equation}
T = mS + \frac{3}{2} K(I)\frac{d{\cD}}{dS} S - 2 K(I){\cD}(S) -
{\cD}\frac{dK}{dI} \Bigl(A_1^2 e^{-2\beta_1} + A_2^2 e^{-2\beta_2} +
A_3^2 e^{-2\beta_3}\Bigr).\label{T}
\end{equation}
In view of \eqref{emtcomp} and \eqref{T} system of Einstein
equations now takes the form
\begin{subequations}
\label{BIDnew}
\begin{eqnarray}
\ddot \alpha - \dot \alpha^2 + \dot \beta_1^2 + \dot \beta_2^2 +
\dot \beta_3^2  &=&  - \frac{\kappa e^{2\alpha}}{2} \Bigl[\Bigl( mS
+ K {\cD} - \frac{3}{2} K \frac{d{\cD}}{dS} S\Bigr) + \Bigl(\dot
A_1^2 e^{-2\alpha} +  {\cD}\frac{dK}{dI}A_1^2 \Bigr)
e^{-2\beta_1}\nonumber\\
&+& \Bigl(\dot A_2^2 e^{-2\alpha} +  {\cD}\frac{dK}{dI} A_2^2 \Bigr)
e^{-2\beta_2} + \Bigl(\dot A_3^2 e^{-2\alpha} +  {\cD}\frac{dK}{dI}
A_3^2 \Bigr) e^{-2\beta_3}\Bigr],\label{00new}\\
\ddot \beta_1 &=&  -\frac{\kappa e^{2\alpha}}{2} \Bigl[ -\Bigl(mS +
\frac{1}{2} K \frac{d{\cD}}{dS} S  - K {\cD}\Bigr) +  \Bigl(\dot
A_1^2 e^{-2\alpha} - {\cD} \frac{dK}{dI} A_1^2\Bigl) e^{-2\beta_1}
\nonumber\\ &-&  \Bigl(\dot A_2^2 e^{-2\alpha} - {\cD} \frac{dK}{dI}
A_2^2\Bigl) e^{-2\beta_2} -  \Bigl(\dot
A_3^2 e^{-2\alpha} - {\cD} \frac{dK}{dI} A_3^2\Bigl) e^{-2\beta_3}\Bigr],\label{11new}\\
\ddot \beta_2 &=&  -\frac{\kappa e^{2\alpha}}{2} \Bigl[ - \Bigl(mS +
\frac{1}{2} K \frac{d{\cD}}{dS} S  - K {\cD}\Bigr) +  \Bigl(\dot
A_2^2 e^{-2\alpha} - {\cD} \frac{dK}{dI} A_2^2\Bigl) e^{-2\beta_2}
\nonumber\\ &-&  \Bigl(\dot A_3^2 e^{-2\alpha} - {\cD} \frac{dK}{dI}
A_3^2\Bigl) e^{-2\beta_3} -  \Bigl(\dot
A_1^2 e^{-2\alpha} - {\cD} \frac{dK}{dI} A_1^2\Bigl) e^{-2\beta_1}\Bigr],\label{22new}\\
\ddot \beta_3 &=&  -\frac{\kappa e^{2\alpha}}{2} \Bigl[ - \Bigl(mS +
\frac{1}{2} K \frac{d{\cD}}{dS} S  - K {\cD}\Bigr) + \Bigl(\dot
A_3^2 e^{-2\alpha} - {\cD} \frac{dK}{dI} A_3^2\Bigl) e^{-2\beta_3}
\nonumber\\ &-&  \Bigl(\dot A_1^2 e^{-2\alpha} - {\cD} \frac{dK}{dI}
A_1^2\Bigl) e^{-2\beta_1} -   \Bigl(\dot A_2^2 e^{-2\alpha} - {\cD}
\frac{dK}{dI} A_2^2\Bigl) e^{-2\beta_2}\Bigr]. \label{33new}
\end{eqnarray}
\end{subequations}

The  triviality of off-diagonal components of the Einstein tensor
for BI metric leads to
\begin{equation}
T_2^1 = T_3^2 = T_1^3 = 0, \label{trivoff}
\end{equation}
from which one finds
\begin{equation}
\frac{\dot A_1}{A_1} \frac{\dot A_2}{A_2} = \frac{\dot A_2}{A_2}
\frac{\dot A_3}{A_3} = \frac{\dot A_3}{A_3} \frac{\dot A_1}{A_1} =
{\cD}\frac{dK}{dI} e^{2\alpha}. \label{trivoffnew}
\end{equation}
From \eqref{trivoffnew} one easily finds
\begin{equation}
\frac{\dot A_1}{A_1} = \frac{\dot A_2}{A_2} = \frac{\dot A_3}{A_3},
\label{rela123}
\end{equation}
leading to the following relations between the three components of
vector potential:
\begin{equation}
A_1 = A, \quad A_2 = C_{21}A, \quad A_3 = C_{31}A,
\label{rela123new}
\end{equation}
with $C_{21}$ and $C_{31}$ being constants of integration.

In view of \eqref{trivoff} and  \eqref{rela123} one easily finds
that
\begin{subequations}
\begin{eqnarray}
\ddot \alpha - \dot \alpha^2 + \dot \beta_1^2 + \dot \beta_2^2 +
\dot \beta_3^2  &=&  - \frac{\kappa e^{2\alpha}}{2} \Bigl[\Bigl( mS
+ K {\cD} - \frac{3}{2} K \frac{d{\cD}}{dS} S\Bigr) - 2{\cD} I
\frac{dK}{dI}
\Bigr],\label{ddotalpha}\\
\ddot \beta_1 = \ddot \beta_2 = \ddot \beta_3 &=& \frac{\kappa
e^{2\alpha}}{2}\Bigl[mS + \frac{1}{2} K \frac{d{\cD}}{dS} S  - K
{\cD}\Bigr]. \label{ddotbeta}
\end{eqnarray}
\end{subequations}
In view of \eqref{rela123} from \eqref{emf123} it can be shown that
\begin{equation}
\dot \beta_1 = \dot \beta_2 = \dot \beta_3. \label{eqdotbets}
\end{equation}
In view of \eqref{ddotbeta} and \eqref{eqdotbets} one concludes that
the $\beta$s differ by some constant only, namely
\begin{equation}
\beta_1 = \beta, \quad \beta_2 = \beta + \beta_{21}, \quad \beta_3 =
\beta + \beta_{31}, \label{beta}
\end{equation}
with $b$, $\beta_{21}$ and $\beta_{31}$ being some arbitrary
constants. Since these constants leads to the different scaling
along different axis, there is only one option left, it is to set
$\beta_{21} = \beta_{31} = 0$. That means the current model allows
isotropization. Inserting ${\cD}\frac{dK}{dI} e^{2\alpha}$ into
\eqref{emf123} one finds the equation for $A$:
\begin{equation}
A \ddot A + \dot A^2 - 2 \dot \beta A \dot A = 0, \label{eqA}
\end{equation}
with the solution
\begin{equation}
A = \sqrt{C_1 \int e^{2\beta} dt + C_2}, \label{Agen}
\end{equation}
where $C_1$ and $C_2$ are arbitrary constants.

On account of \eqref{rela123new} and \eqref{beta} one now finds
\begin{equation}
I = - Q A^2 e^{-2 \beta}, \quad Q = 1 + C_{21}^{2}  + C_{31}^{2} =
{\rm const.} \label{Invnew}
\end{equation}
In view of \eqref{cc}, \eqref{ddotbeta} and \eqref{eqdotbets} the
equation \eqref{ddotalpha} can be rearranged as
\begin{equation}
6 \dot \beta^2 = \kappa \Bigl[2mS - {\cD}
\frac{d}{dI}\Bigl(IK\Bigr)\Bigr] e^{2\alpha}. \label{alb}
\end{equation}
In what follows, we consider some some concrete cases.

{\bf Massless spinor field with Heisenberg-Ivanenko nonlinearity}

Let us consider the massless spinor field with Heisenberg-Ivanenko
nonlinearity. Note that, in the unified nonlinear spinor theory of
Heisenberg, the massive term remains absent, and according to
Heisenberg, the particle mass should be obtained as a result of
quantization of spinor prematter~ \cite{massless}. In the nonlinear
generalization of classical field equations, the massive term does
not possess the significance that it possesses in the linear one, as
it by no means defines total energy (or mass) of the nonlinear field
system. Thus without losing the generality we can consider the
massless spinor field putting $m\,=\,0.$ For Heisenberg-Ivanenko
nonlinearity we have ${\cD}(S) = \sigma S^2$. In this case Eq.
\eqref{ddotbeta} takes the form
\begin{equation}
\ddot \beta = 0, \label{betaHI} \end{equation}
with the solution
\begin{equation}
\beta = bt + b_1, \label{metf}
\end{equation}
where $b$ and $b_1$ are arbitrary constants. In view of \eqref{metf}
Eq. \eqref{eqA} now reads
\begin{equation}
A \ddot A + \dot A^2 - 2 b A \dot A = 0. \label{eqA}
\end{equation}
The equation \eqref{eqA} allows the following solution:
\begin{eqnarray}
A = \sqrt{e^{2bt} - C}, \label{sqA}
\end{eqnarray}
Here $C$ is an arbitrary constant. In what follows, we consider a
specific solution to the Eq. \eqref{eqA}:
\begin{eqnarray}
A = D e^{bt}, \label{A}
\end{eqnarray}
with $D$ being an arbitrary constant. Inserting \eqref{A} into
\eqref{Invnew} one finds
\begin{equation}
I = - \bar{Q} D^2, \quad \bar{Q} = e^{2b_1} Q = e^{2b_1}[1 +
C_{21}^{2} + C_{31}^{2}]. \label{Invnew1}
\end{equation}

On account of \eqref{A} and ${\cD}$ from the triviality off-diagonal
components of energy-momentum tensor we find
\begin{equation}
K = \frac{b^2}{\sigma S_0^2} I + C_3, \label{K}
\end{equation}
with $C_3$ being some arbitrary constant. Finally inserting $K$ from
\eqref{K} into \eqref{alb} one finds

\begin{equation}
(6 + 2 \kappa \bar{Q} D^2) b^2  = \kappa \sigma S_0^2 C_3.
\label{albcons}
\end{equation}

Equation \eqref{albcons} gives the relation between different
constants.

{\bf Case with minimal coupling}

Let us consider the case with minimal coupling. In this case from
the Off-diagonal components of energy-momentum tensor we find
\begin{equation}
\dot A_1 \dot A_2 = \dot A_2 \dot A_3 = \dot A_3 \dot A_1 = 0.
\label{a123}
\end{equation}
From \eqref{a123} follows that at least two of the three components
$A_i$ are constant, which means only one of the components of
$F_{\mu\nu}$ is nonzero. Let us assume that $\dot A_1 = \dot A \ne
0$. In view of $\dot A_2 = \dot A_3 = 0$ from the electromagnetic
field equations in this case we have
\begin{equation}
A = C \int e^{2\beta_1} dt + C_1, \quad A_2 = {\rm const.}, \quad
A_3 = {\rm const.}, \label{emc}
\end{equation}
with $C$ and $C_1$ being some arbitrary constants. Components of the
energy-momentum tensor in this case read
\begin{equation}
T_0^0 - mS  = T_1^1 = - T_2^2 =  - T_3^3 = \frac{C_1^2}{2}
e^{2\beta_1 - 2 \alpha}. \label{emtemf}
\end{equation}
Einstein field equations in this case takes the form
\begin{subequations}
\label{minc}
\begin{eqnarray}
\ddot \alpha - \dot \alpha^2 + \dot \beta_1^2 + \dot \beta_2^2 +
\dot \beta_3^2  &=& -\frac{\varkappa C_1^2}{2} e^{2\beta_1} -
\frac{m \varkappa S_0}{2} e^{\alpha}, \label{00new10}\\
\ddot \beta_1 &=& -\frac{\varkappa C_1^2}{2} e^{2\beta_1} +
\frac{m \varkappa S_0}{2} e^{\alpha}, \label{11new10}\\
\ddot \beta_2 &=& \frac{\varkappa C_1^2}{2} e^{2\beta_1} +
\frac{m \varkappa S_0}{2} e^{\alpha}, \label{11new20}\\
\ddot \beta_3 &=& \frac{\varkappa C_1^2}{2} e^{2\beta_1} + \frac{m
\varkappa S_0}{2} e^{\alpha}. \label{11new30}
\end{eqnarray}
\end{subequations}
Unlike the case with interacting electromagnetic and scalar fields,
in case of minimal coupling we have $\ddot \beta_1 \ne \ddot \beta_2
= \ddot \beta_3$, which shows the space-time in this case
essentially anisotropic. Addition of \eqref{11new10},
\eqref{11new20} and \eqref{11new30}, on account of coordinate
condition, gives
\begin{equation}
\ddot \alpha =  \frac{\varkappa C_1^2}{2} e^{2\beta_1} + \frac{3m
\varkappa S_0}{2} e^{\alpha}. \label{11newal0}
\end{equation}
In case of massless spinor field the system \eqref{minc} can be
easily solved to obtain
\begin{equation}
e^{2\beta_1} = - \frac{2\eta^2}{\varkappa C_1^2} {\rm cosech}^2
(\eta t), \quad \eta^2 = {\rm const.}, \label{beta_10}
\end{equation}
and
\begin{equation}
e^{2\alpha} = e^{2\beta_2} = e^{2\beta_3} = \sinh^2{(\eta t)}.
\label{ab2b3}
\end{equation}
Thus we see that direct interaction is essential for isotropization
process of initially anisotropic space-time.

\section{Conclusion}
Within the framework of Bianchi type-I cosmological model evolution
of the initially anisotropic space-time in presence of an
interacting system of spinor and electromagnetic fields is studied.
It is shown that the interacting term can be viewed as effective
photon mass. The present model allows asymptotic isotropization of
initially anisotropic space-time.


\begin{thebibliography}{99}
\bibitem{gold} A.S. Goldhaber and M.M. Nieto,  Rev. Mod. Phys. {\bf 43},
277 (1971)
\bibitem{froome} K.D. Froome and L. Essen, {\it The velocity of light and radio waves}
Academis, NY (1969).
\bibitem{taylor} B.N. Taylor, H.W. Parker, and D.N. Langenberg, Rev.
M od. Phys. {\bf 41}, 375 (1969).
\bibitem{broglie} L. de Broglie, {\it Theorie generale des particles
a spin} Gauthier-Villars, Paris, 2 nd Ed. (1954).
\bibitem{rosen} A.H. Rosenfeld {\it et al} Rev. Mod. Phys. {\bf 40},
77 (1968).
\bibitem{williams} E.R. Williams, J.E. Faller, and H. Hill, Phys.
Rev. Lett. {\bf 26}, 721 (1971)
\bibitem{crandal} R.E. Crandall, Am. J. Phys. {\bf 51}, 698 (1983).
\bibitem{chernikov} M.A. Chernikov, C.J. Gerber, H.R. Ott, and H.J.
Gerber, Phys. Rev. Lett. {\bf 68}, 3383 (1992).
\bibitem{schaefer} B.E. Schaefer, Phys. Rev. Lett. {\bf 82}, 4964 (1999).
\bibitem{fishbach} E. Fishbach {\it et al} Phys. Rev. Lett. {\bf 73}, 514 (1994).
\bibitem{davis} L. Davis, A.S. Goldhaber, and M.M. Nieto, Phys. Rev. Lett. {\bf 35}, 1402 (1975).
\bibitem{lakes} R. Lakes, Phys. Rev. Lett. {\bf 80}, 1826 (1998).
\bibitem{luo} J. Luo, L.-C. Tu, Z.-K. Hu, and E.-J. Luan, Phys. Rev. Lett. {\bf 90}, 081801 (2003).
\bibitem{shikin1} Yu.A. Popov, Yu.P. Rybakov, G.N. Shikin, and Bijan
Saha {\it Electromagnetic field with induced massive term: Case with
spinor field} ArXiv:
\bibitem{shikin2} B. Saha and G.N. Shikin, J. Math. Phys. {\bf 38}, 5305 (1997).
\bibitem{saha1} Bijan Saha, Phys. Rev. D {\bf 64}, 123501 (2001).
\bibitem{saha2} Bijan Saha, Physics of Particles and Nuclei {\bf 37} Suppl. 1,
S13, (2006).
\bibitem{saha3} Bijan Saha, Phys. Rev. D {\bf 74}, 124030 (2006)
\bibitem{saha4} Bijan Saha,  Physics of Particles and Nuclei {\bf 40}(5),656
(2009).
\bibitem{massless} W. Heisenberg, {\it Introduction to the unified
field theory of elementary particles} (Interscience Publ., London.
1966).
\end{thebibliography}
\end{document}